 \documentclass[prd,preprintnumbers,amsmath,amssymb,floatfix]{revtex4}

\usepackage{graphicx}
\usepackage{dcolumn}
\usepackage{bm}
\usepackage{amssymb}
\usepackage{epsfig}
\usepackage{color}

\newcommand{\ba}{\begin{eqnarray}}
\newcommand{\ea}{\end{eqnarray}}
\newcommand{\be}{\begin{equation}}
\newcommand{\ee}{\end{equation}}
\newcommand{\bdisplay}{\begin{displaymath}}
\newcommand{\edisplay}{\end{displaymath}}

\newcommand{\eq}[1]{Eq.\,(\ref{#1})}

\newcommand{\sigin}{\sigma_{\rm inel}}
\newcommand{\sigel}{\sigma_{\rm el}}
\newcommand{\sigtot}{\sigma_{\rm tot}}
\begin{document}

\title{``Soft" Hadronic Cross Sections Challenge Hidden Dimensions}  
\author{Martin~M.~Block}
\affiliation{Department of Physics and Astronomy, Northwestern University, 
Evanston, IL 60208}
\author{Francis Halzen}
\affiliation{Department of Physics, University of Wisconsin, Madison, WI 53706}
\date{\today}

\begin{abstract}
High energy measurements of the inelastic proton-proton cross sections, at the LHC at $\sqrt s$=7 TeV and by  Auger at 57 TeV, have validated previous evidence from data collected over a wide range of energies that the total and inelastic cross sections for $pp$ and $\bar pp$ interactions saturate the Froissart bound of $\ln^2 s$. 
Although the data themselves did not cover truly asymptotic energies, our recent analysis of these data obtained the  asymptotic ratio $\sigin/\sigtot=0.509\pm 0.021$, consistent with the value of 1/2 required for scattering by a black disk; further, the forward scattering amplitude became purely imaginary for $s\rightarrow \infty$, confirming the black disk interpretation. In addition, the limiting black disk behavior has been independently confirmed  by an  analysis of Schegelsky and Ryskin  including LHC data on the shrinkage of the slope of the forward elastic scattering cross section. Unless one considers these results, emerging from an analytic amplitude analysis of data over an energy range of $6\le \sqrt s\le 57000$ GeV, a complete numerical  accident, we rule out any new physics thresholds that contribute higher powers of $\ln s$ or, worse, powers of $s$ to the energy dependence of cross sections {\it at any energy}. This includes theories with additional dimensions of space-time, whose existence is challenged.
\end{abstract}
\maketitle
\section{Introduction} The idea that space may have additional hidden dimensions has been through many reincarnations: from Kaluza who extended general relativity to 5 dimensions, to string theories for modeling the Pomeron, to, finally, their reinterpretation as quantum theories of gravity.  It has also been suggested that the unification of gravity with the Standard Model interactions may occur at LHC energies, implying the existence of sub-millimeter size dimensions of space. We will here revisit the fact that the thresholds associated with the appearance of additional dimensions leave an imprint on the energy dependence of ``soft" cross sections, specifically of the forward elastic scattering amplitudes that determine the total, elastic and inelastic hadronic cross sections. Subsequently, we will show that there is experimental evidence against such thresholds in $s$,  where $s$ is the square of the center of mass energy (cms), not only at LHC ($\sqrt s= 7$ TeV)  and cosmic ray energies ($\sqrt s= 57$ TeV), but, tantalizingly, to {\em infinite energy}.

Convincing experimental evidence has emerged for a picture where hadrons are asymptotically black disks  \cite{disk}, presumably of gluons, whose radius grows as $\ln s$. The energy dependence of total cross sections is indeed dominated by a $\ln^2 s$ term, a fact that first emerged from a determination of forward $pp$ and $\bar pp$ scattering amplitudes constrained by analyticity and anchored to data, from energies as low as 6\,GeV, where they are constrained by finite energy sum rules that exploit very accurate low energy data, to measurements at the Tevatron. Subsequently, this result has been quantitatively confirmed by measurements of the total and inelastic $pp$ cross section at the LHC and, at even higher energy, by the Pierre Auger Observatory (PAO) air shower array. Without imposing a priori constraints on their asymptotic behavior, we have found that other required properties of a black disk emerged from the extrapolation of the forward scattering amplitudes to infinite energy : a purely imaginary amplitude and a ratio of the inelastic to the total cross sections $\sigin/\sigtot$ consistent with $1/2$ within a small statistical error. The amplitudes also correctly anticipated the shrinkage of the high energy forward elastic scattering cross section; it has been pointed out in Ref. \cite{Ryskin} that their measurement at the LHC provides totally independent evidence for the proton as an asymptotic black disk.

The conclusion of this note will be that there is no room for hidden dimensions, from presently available energies to infinite energy, because their thresholds contribute additional logarithms or, worse, powers of s to the energy dependence of the cross section \cite {Yogi}. These additional contributions to the energy dependence {\em cannot be hidden}. While challenging the validity of string theories, our analysis also opens the prospect of probing fundamental physics questions of analyticity and unitarity by exploring soft hadron measurements using accelerator and cosmic ray beams.

In this note, we first recall how hidden dimensions contribute to forward hadronic scattering. We next itemize the varied and increasingly quantitative pieces of evidence for an asymptotic black disk structure of hadrons, confronting our high energy predictions for $\sigtot$ and $\sigin$ with recent $pp$  LHC and PAO measurements. We conclude, unlike in Ref. \cite{Yogi}, that there is no room at any energy  for additional thresholds associated with the appearance of hidden dimensions.
\section{The Signature of Hidden Dimensions in Soft Hadron Scattering}
As we will discuss further on, there is compelling evidence that the total scattering cross section, e.g., $\sigma_{\rm tot} (pp)$ for the scattering of protons, increases with center of mass energy $E$ as $\ln^2 E$, a dependence often associated with Froissart's $\ln^2s$ unitarity bound \cite{froissart}; $E={\sqrt s}/2$. A heuristic way to visualize this energy dependence follows from considering the elastic scattering of two particles, dominated by the exchange of the lightest particle of mass M, with a probability
\ba
P(r,E) \sim \exp\,(-2Mr+ \frac {S(E)}{k}),\label{P}
\ea
where $S$ is the entropy that counts the number of final states and $k$ is Boltzmann's constant;
it becomes of order unity for a range $R$, where 
\ba
 R&=&\frac {S(E)}{2kM}. \label{R}
\ea The total cross section for shadow scattering is
\begin{eqnarray}
\sigtot=2 \pi  R^2 = \frac {\pi}{2M^2}\,\, [\frac {S}{k}]^2.\label{sigtot1}
\end{eqnarray}
The unitarity bound corresponds to a growth of entropy with energy as $S(E) \simeq \ln E$. Alternatively, one can think of the final states in terms of an ensemble of relativistic particles, or, from the microscopic point of view, the growth with energy of the number of cells in the ladder diagrams of the multiperipheral model. From the parton point of view, the picture that emerges asymptotically is that of a proton composed of an increasing number of soft gluon constituents, each carrying a decreasing fraction of the proton energy. All lead to a $\ln s$ increase of the entropy and a $\ln^2 s$ rise of the cross section.  Where $M$ in \eq{sigtot1}  was historically identified with the mass of the pion, it is now associated with the particles populating the Pomeron trajectory, i.e., glueballs.

The value of this heuristic presentation of a cross section behaving as $\ln^2 E$  is that it readily demonstrates that this energy dependence is untenable in the presence of additional dimensions of space-time. In a string picture, the gravitational force is represented by a stretching string with tension $\tau_G \sim 1/G$; here G is Newton's constant. In theories with hidden compact dimensions of size $\Lambda$, the string tension is reduced to $\tau_G/\Lambda^n$, where n is the number of additional dimensions of space. Diluting the tension by additional dimensions is the origin of the suppression of the strength of gravity relative to the Standard Model interactions. The extra degrees of freedom n, with $n\ge1$, introduce an {\it  additional} contribution $S_n$ to the $\ln E$ term in the entropy of Eq.\,2 given by
\ba
\frac {S_n(E)}{k} \simeq \sqrt{\frac {\Lambda^n}{\tau_G}} E^{1+\frac{n}{2}}.\label{Sn}
\ea
Using \eq{R} and \eq{Sn} in \eq{sigtot1}, we conclude that additional dimensions violate the $\ln^2 s$ previously obtained, with a power-law contribution that dominates above an energy threshold $s_0$ fixed by the size ${\cal R} \sim \Lambda^{-1}$ of the additional dimensions (the smaller the size of the hidden dimension the larger the effective threshold $s_0$), so for $s\gg s_0$,
\ba
\sigtot \sim G \,\Lambda^n\, s^{1+\frac {n}{2}}.\label{sigtot}
\ea
The contribution of the additional dimensions to the entropy and therefore to the cross section behaves as a power law in $s$, and therefore dominates after crossing the energy threshold $s_0$ determined by the size of the additional dimension; below this threshold the cross section reverts to a $\ln^2 s$ energy dependence. We have here sketched the treatment of Ref.  \cite{Yogi}. A similar result is generically obtained in string theories with only the coefficients of the logarithmic and power-behaved terms varying according to the specific model. 

It may be counterintuitive, but this result implies that additional dimensions should be directly visible in measurements of soft hadronic cross sections, as well as in the more familiar highly dedicated analyses required to expose their signature in the presence of large backgrounds. As we will argue next, over time evidence has accumulated, quantitatively confirmed by recent LHC and cosmic ray data, that the cross sections not only behave as $\ln^2 s$ at LHC  and cosmic ray energies, but that this behavior persists at all energies, leaving no room for hidden dimensions at any energy, unlike the conclusion of Ref. \cite{Yogi}, since we show the evidence for the {\em asymptotic black disk proton} in the next Section.

\section{Hadrons Are Asymptotically Black Disks: the Evidence}

\begin{figure}[h] 
\begin{center}
\mbox{\epsfig{file=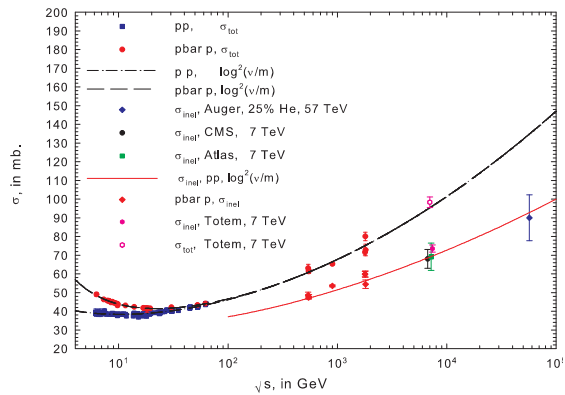
,width=3in%
,bbllx=150pt,bblly=375pt,bburx=595pt,bbury=680pt,clip=%
}}
\end{center}
\caption[]{
The upper curves show the fitted total cross section, $\sigtot$,  for $\bar pp$ (dashed curve)  and $pp$ (dot-dashed curve), in mb vs. $\sqrt s$, the cms energy in GeV, taken from Ref. \cite{BH}. The $\bar pp$ data used in the fit  are the (red) circles and the $pp$ data are the (blue) squares. The fitted data were anchored by values of $\sigtot^{\bar pp}$ and $\sigtot^{pp}$, together with the energy derivatives  ${d\sigtot^{\bar pp}/ d\nu}$ and ${d\sigtot^{pp}/ d\nu}$ at 4 GeV using FESR \cite{FESR}, as described in Ref. \cite{BH,physicsreports}. The (pink) open hexagon, the Totem total cross section \cite{Totemsigma} at 7 TeV, was not used in the fit, but is displayed for comparison.   
The lower  (red) solid curve taken from Ref. \cite{disk} is our {\em predicted} inelastic cross section, $\sigin$, in mb,  vs. $\sqrt s$, in GeV. The lowest energy inelastic data, the  $\bar pp$ (red) diamonds, were {\em not} used in the fit, nor were the 3 LHC high energy $pp$ inelastic measurements, the (black) circle  CMS \cite{CMS}, the (green) square Atlas \cite{Atlas}  and the (pink) hexagon Totem measurements \cite{Totemsigma} at 7 TeV, as well as  the (blue) diamond PAO measurement \cite{Auger} at 57 TeV. As clearly seen, our asymptotic  $\ln^2s$ predictions both for the total cross section at 7 TeV and the inelastic cross sections at 7 and 57 TeV are in excellent agreement with the very high energy measurements.\label{fig:pp}
}
\end{figure}

We next collect the accumulating, and increasingly compelling, experimental evidence that particle cross sections saturate the Froissart bound, leaving no room for additional contributions in the form of powers, or even logarithms of $s$, up to $\sqrt s=57$ TeV. Our analysis in Ref. \cite{disk} of this data, constrained by analyticity, did more: it delineated the $\ln^2 s$ behavior of the cross sections up to infinity. Let us itemize the evidence:

\begin{enumerate}
\item Using a parameterization of the forward scattering amplitudes constrained by analyticity \cite{BH}, we have argued for some time that data on $\bar pp$ and $pp$ total cross section show evidence for saturation of the Froissart bound, i.e., fits to their energy dependence pinpoints a $\ln^2 s$ high energy behavior. The evidence became most compelling when this description of the lower energy data quantitatively anticipated measurements at the LHC and by the PAO air shower array. These amplitudes now describe the data over an energy range from $6\le \sqrt s \le 57000$ GeV; at energies below 6 GeV they satisfy finite energy sum rules (FESR) \cite{FESR} that fix both the total cross sections and their derivatives for both $\bar pp$ and $pp$. The analysis yields an even amplitude cross section $\sigma^0$, which describes the common high energy behavior for both $\bar pp$ and $pp$, and is   given by
\ba
\sigma^0(\nu)&\equiv&\beta_{\cal P'}\left(\frac{\nu}{m}\right)^{\mu -1}+c_0+c_1\ln\left(\frac{\nu}{m}\right)+c_2\ln^2\left(\frac{\nu}{m}\right),\label{sig0pp}
\ea 
where $\nu$ is the laboratory energy of the incoming proton (anti-proton), $m$ the proton mass, and the `Regge intercept' $\mu=0.5$. The predictions for the $pp$ and $\bar pp$ total cross sections are shown as the upper curve(s) in Fig. \ref{fig:pp}, whose parameters are given in Table \ref{tab:sigma0}.
\begin{table}[h,t]                   
%
\def\arraystretch{1.5}            

\begin{center}
\caption[]{Values of the parameters, taken from Ref. 
\cite{BH}, for the even amplitude, $\sigma^0(\nu)$, using 4 FESR analyticity constraints. 
\label{tab:sigma0}
}
\vspace{.2in}
\begin{tabular}[b]{||l||l||l||l||}
\hline\hline
$c_0$=$37.32$ mb,
&$c_1$=$-1.440\pm 0.070$ mb,
&$c_2$=$0.2817\pm 0.0064$ mb,
&$\beta_{\cal P'}$=$37.10$ mb\\
\hline\hline
\end{tabular}
\end{center}
\end{table}
\def\arraystretch{1}  
\item The same approach \cite{BH} revealed a forward scattering amplitude that is purely imaginary as $s\rightarrow\infty$.
\item Incorporating these amplitudes into an eikonal description of the nucleon \cite{physicsreports}, we obtained predictions for the ratio of the elastic to total cross section $r= \sigin/\sigtot$, as well as the slope parameter $B\equiv [\frac{d}{dt}\ln d\sigel/dt]_{t=0}$, the logarithmic derivative of the forward  differential elastic scattering cross section, where $t$ is the square of the 4-momentum transfer. 
\item At this stage, our analysis allowed us to predict the inelastic cross section $\sigma_{\rm inel}^0$ \cite{disk}. To obtain its most accurate value,  the total cross section of \eq{sig0pp} was multiplied by $r$ to obtain an inelastic cross section,  $\sigma^0_{\rm inel}(\nu)$, given by
\ba
\sigma^0_{\rm inel}(\nu)&\equiv&\beta_{\cal P'}^{\rm inel}\left(\frac{\nu}{m}\right)^{\mu -1}+c_0^{\rm inel}+c_1^{\rm inel}\ln\left(\frac{\nu}{m}\right)+c_2^{\rm inel}\ln^2\left(\frac{\nu}{m}\right)\label{cinel}\\
&=& 62.59\left(\frac{\nu}{m}\right)^{-0.5}+24.09+0.1604 \ln\left(\frac{\nu}{m}\right)+ 0.1433 \ln^2\left(\frac{\nu}{m}\right) \ {\rm mb}.   
\label{finalinelastic}
\ea 
 Fig.\,\ref{fig:pp} shows the prediction for the inelastic cross section from \eq{finalinelastic} as the lower solid (red) curve. The experimental data shown are {\em not} used in the fit, but are displayed for comparison.  Obviously, they are in excellent agreement with the predicted inelastic cross section. 
\item  This agreement yielded a key clue: the {\em asymptotic} value for  $\sigin/\sigtot$ is given by the ratio of the $\ln^2 s$ coefficients of \eq{sig0pp}, \eq{cinel} and \eq{finalinelastic},  $c_2^{\rm inel}/c_2=0.509\pm 0.021$, compatible with the black disk value of 1/2. This allowed us  \cite{disk} to conclude that the proton becomes a black disk as $s\rightarrow \infty$. However, if the total  cross section had contained a power of $s$, such as the power law of \eq{sigtot} resulting from hidden dimensions appearing above a threshold energy $s_0$, this power law term would dominate  as $s\rightarrow \infty$ and the ratio of the (now) lower order terms $c_2^{\rm inel}/c_2$ would be numerically meaningless as an asymptotic limit. 
\item The total cross section for a black disk is given by $\sigma_{\rm tot}=2\pi R^2$, where $R=R_0\ln s$. Since  $B=R^2/4$, the nuclear slope parameter $B$ measures its energy dependent radius $R$,  so that $\sigma_{\rm tot}=8\pi B$ . A fit to $B$ by Schegelsky and Ryskin \cite{Ryskin} which included the recent measurement at the LHC by the TOTEM collaboration \cite {Totemdsigma} not only quantitatively confirmed our predictions \cite{disk}, but in so doing, provided a totally independent verification  for the saturation of the $\ln^2s$ Froissart bound. Again, this agreement would be violated if there were additional hidden dimensions. 
\end{enumerate}
 In a parton picture a black disk is not unexpected; with increasing energy the proton is populated by a growing number of gluons, each carrying a progressively smaller fraction of the proton momentum. The proton becomes an absorbing disk of gluons, as do all other hadrons as well as the photon, through vector meson dominance. 
Asymptotically all hadrons are identical because quarks, and therefore the hadron's quantum numbers, no longer play a role. This picture, anticipated by our analytic amplitude analysis performed up to Tevatron energy, has been validated by all subsequent measurements of $\sigtot, \sigin$ and $B$ at the LHC and by cosmic ray experiments. Within the partonic interpretation of our analytic amplitudes, yet one more prediction emerges: if one ascribes the origin of the asymptotic $\ln^2 s$ term in $\bar p p$ and $pp$ scattering to gluons only, then it is universal and its energy dependence as well as its normalization is  the same for $\pi \pi$, $\pi p$, $Kp$, and $\gamma p$ interactions via vector meson dominance. This is indeed the case; see Ref. \cite{glue} for an update.

\section{Conclusion}

Unless one considers the convergence of cross section data on an asymptotic  black disk structure of hadrons a total accident, one must conclude that there is no room for new physics contributing additional logarithms or powers of $s$ to cross sections, not only at LHC energy, but at all energies. Importantly, as argued at the beginning of this paper (see Ref. \cite{Yogi}), this excludes additional dimensions of space-time. No matter how high in energy their threshold, they would have asymptotically destroyed the by now highly constrained properties of the black disk.

{\section{Acknowledgments} 
In part,  F. H. is supported by the National Science Foundation Grants OPP-0236449 and PHY-0969061, by the DOE grant DE-FG02-95ER40896 and  by the University of Wisconsin Alumni Research Foundation. M. M. B. thanks  the Aspen Center for Physics, supported in part by NSF Grant No. 1066293,  for its hospitality during this work and also thanks Prof. D. W. McKay for valuable conversations. 

\end{document}